%% file: main.tex
\documentclass[aps,pre,reprint,floatfix,longbibliography,superscriptaddress]{revtex4-1}
\usepackage{graphicx} 
\usepackage{amssymb,amsfonts,amsmath}
\usepackage{braket} \usepackage{color} 
\usepackage{multirow}
\usepackage{dcolumn}
\usepackage[mathlines]{lineno}
\usepackage[compat=1.1.0]{tikz-feynman}
\begin{document}

\title{Phonon Quantum Phase Transition}
\input{authors}

%\date{\today}

\begin{abstract}
We show that a quantum phase transition can occur in a phonon system in the presence of dislocations. Due to the competing nature between the topological protection of the dislocation and anharmonicity, phonons can reach a quantum critical point at a frequency determined by dislocation density and the anharmonic constant, at zero temperature. In the symmetry-broken phase, a novel phonon state is developed with a dynamically-induced dipole field. We carry out a renormalization group analysis and show that the phonon critical behavior differs wildly from any electronic system. In particular, at the critical point, a single phonon mode dominates the density of states and develops an exotic logarithmic divergence in thermal conductivity. This phonon quantum criticality provides a completely new avenue to tailor phonon transport at the single-mode level without using phononic crystals. 
\end{abstract}

\maketitle

Quantum criticality underlies many unique properties observed in complex quantum systems. Notably, it can promote the emergence of new phases at the onset of a quantum phase transition (QPT) \cite{sachdev2011quantum}, which results from competing ground states separated by a critical value of a non-thermal order parameter, such as pressure or magnetic field strength \cite{vojta2003quantum}. QPTs have been extensively investigated, both experimentally and theoretically, in a host of strongly correlated electron systems, prominently in heavy fermion compounds \cite{gegenwart2008quantum, custers2003break, park2006hidden, si2010heavy, nakatsuji2008superconductivity, gegenwart2007multiple, si2001locally} and high-temperature superconductors \cite{sachdev2000quantum,senthil2004deconfined,van2003quantum,sachdev2008quantum,cooper2009anomalous,shibauchi2014quantum,dai2015antiferromagnetic,sandvik2007evidence}. However, despite scattered reports of QPTs in Bosonic systems such as liquid helium~\cite{fisher1989boson}, trapped ultracold atoms \cite{greiner2002quantum}, and photons \cite{greentree2006quantum}, a comparable study of quantum criticality in phonon systems remains absent, even with the prevalence of phonon-mediated processes and interactions in materials.

In this Letter, we show that phonon systems can exhibit quantum critical phenomena which may be harnessed to evoke novel phonon states with wildly different transport properties compared to conventional phonon systems, and with scaling behavior that is quite distinct from any electronic system. In particular, since phonons typically coexist with impurities and extended defects in a rich interacting environment, we consider competing interactions between phonons and dislocations as a mechanism for emergence of a phonon quantum critical point at $0$K. Moreover, a phonon QPT offers an entirely new way to tune phonon transport by allowing a single phonon mode to dominate the density of states. This is significant because the broadband nature of phonons poses a considerable challenge for controllable phonon transport. Recent efforts have focused on using nanostructured phononic crystals to control phonon transport by engineering the phonon bandstructure \cite{seif2018thermal, xie2018ultra} but require complex artificial structures and are effective only in the sub-THz range. Our work proposes an alternate route to tune phonon transport which may remedy these existing issues, and on a fundamental level may lead to the discovery of exotic phonon phases.

\begin{figure}
  \centering
  \includegraphics[width=8cm]{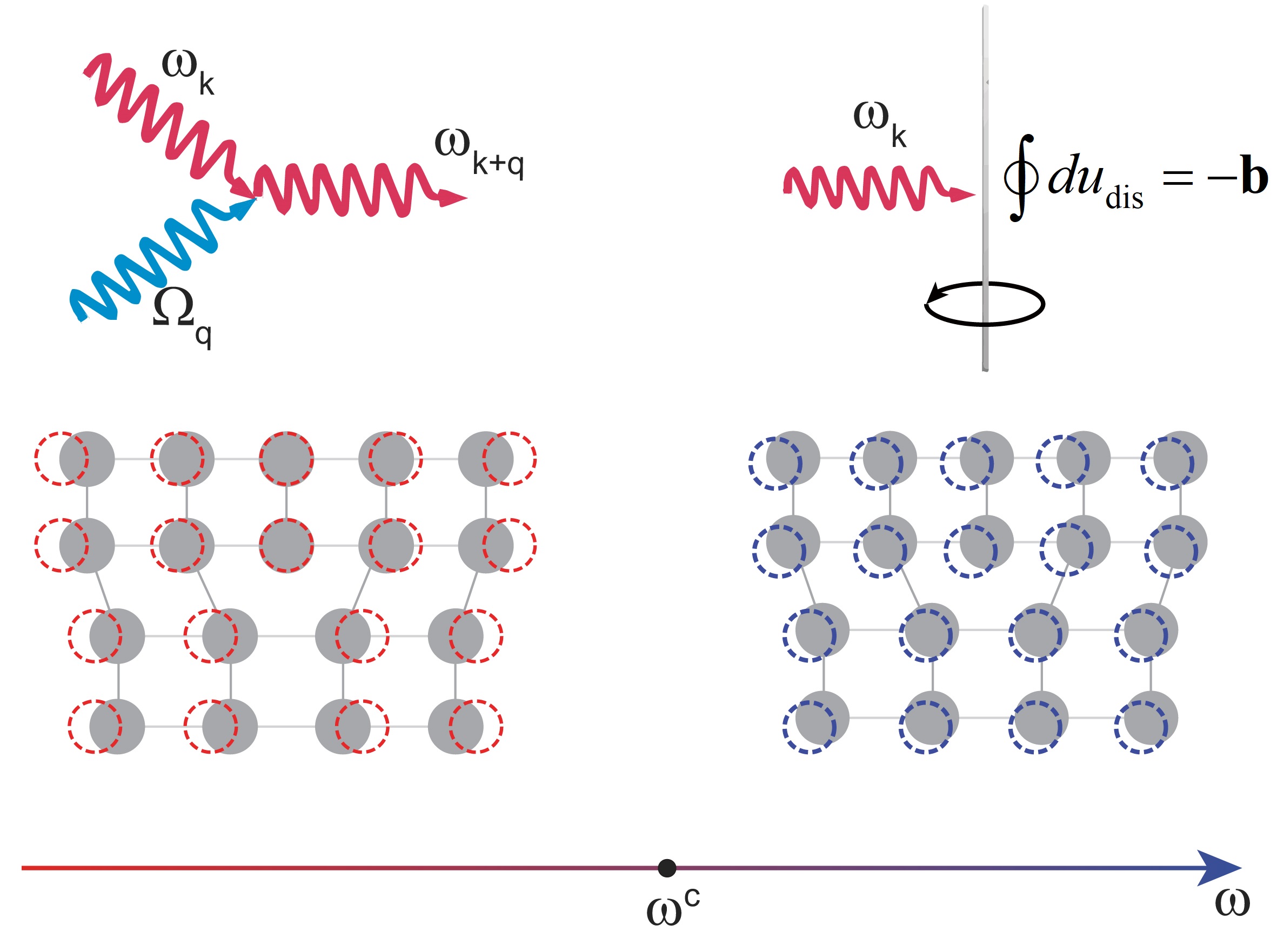}
  \caption{The schematic of a phonon quantum phase transition, caused by competing phonon-dislocation anharmonic interaction (left) and topological protection (right), and separated by a critical point $\omega^{c}$.}
  \label{figureone}
   \end{figure}

When phonons and dislocations coexist in a crystal, their interaction results in phonon scattering by both the static strain field of the dislocation \cite{klemens1955scattering,carruthers1961theory} and dynamic vibration of the dislocation line \cite{ninomiya1968dislocation}, see Fig.~\ref{figureone}. We adopt the recently developed quantized dislocation formalism \cite{li2017nonperturbative,li2018theory} to facilitate the study of phonon-dislocation interaction using a functional integral approach. Considering only the phonon component that couples to a dislocation, the non-interacting phonon action in terms of the phonon displacement field $\phi$ is given by
\begin{equation}
S_\text{ph}[\phi]=\sum_{n,\textbf{k}}\overline{\phi}_{n\textbf{k}}(-i \omega_n + \omega_\textbf{k})\phi_{n\textbf{k}}
\end{equation}
where $\omega_n = 2\pi n /\beta$ are the Bosonic Matsubara frequencies with $\beta \equiv 1/T$, $\overline{\phi}_{n\textbf{k}}=\phi_{n\textbf{k}}^{*}=\phi_{-n-\textbf{k}}$ is the conjugate field, and $\omega_\textbf{k}$ is the phonon dispersion relation. Similarly, the part of the dislocation action that couples with a phonon can be written as~\cite{li2018theory} 
\begin{equation}
S_{\text{dis}}[d] = \sum_{n, \textbf{k} \geq 0} \overline{d}_{n\textbf{k}}(-i\omega_n +\Omega_\textbf{k})d_{n\textbf{k}}
\label{eqn2}
\end{equation}
which is subject to a boundary condition $\lim_{k_{z} \rightarrow 0} d_\textbf{k} \equiv d_{\textbf{s}0} = \sqrt{\beta}C_\textbf{s}$ originating from the topological nature of the dislocation, in which $\textbf{s}$ is the wavevector perpendicular to the dislocation direction, $C_\textbf{s}$ depends on intrinsic parameters of the dislocation defined in Ref.~\cite{li2018theory}, and $\Omega_\textbf{k}$ is the dispersion relation of a dynamic vibrating dislocation.

To compute the third-order (anharmonic) phonon-dislocation interaction, we consider the total lattice displacement $u_\text{tot}=u_\text{dis}+u_\text{ph}$ as a sum over the phonon $u_\text{ph}$ and dislocation $u_\text{dis}$ contributions. We approximate the phonon-dislocation anharmoninc interaction by its dominant contribution $H_\text{anh} \sim u_\text{dis}u_\text{ph}^2$, corresponding to the most common scenario, in which one incoming phonon is scattered into another through a single dislocation. This anharmonic Hamiltonian can then be rewritten in action form as 
\begin{equation}
S_\text{anh}[\phi,d] = \frac{n_\text{dis}}{\sqrt{\beta}}\sum_{n,\textbf{k}\geq 0}(\Phi_{n\textbf{k}}d_{n\textbf{k}}+\Phi_{-n-\textbf{k}}\overline{d}_{n\textbf{k}})
\end{equation}
where   $\Phi_{n\textbf{k}}= \sqrt{L}\sum_{p,q,\textbf{k}_1,\textbf{k}_2}A_\textbf{k}\delta^3(\textbf{k}_1+\textbf{k}_2+\textbf{k}) \delta_{p+q+n,0}\\ \times \phi_{p\textbf{k}_1}\phi_{q\textbf{k}_2}$ is a composite operator of phonon fields $\phi$, and $A_{\textbf{k}}$ is the phonon-dislocation anharmonic coupling strength. The $n_\text{dis}$ prefactor originates from an independent dislocation assumption, where one dislocation is present in a system of dimension $L$, and thus gives the dislocation density $n_\text{dis}=1/L^2$. The total phonon effective action $S[\phi]$ can then be defined as
\begin{equation}
\begin{aligned}
e^{-S[\phi]} &= e^{-S_\text{ph}[\phi]}\int D[d]e^{-S_\text{dis}[d]-S_\text{anh}[\phi,d]} \\
& \times \prod_{n, \textbf{s} \geq 0} \delta(d_{n\textbf{s}0}-\sqrt{\beta}C_\textbf{s})\delta(\overline{d}_{n\textbf{s}0}-\sqrt{\beta}C_\textbf{s})
\end{aligned}
\end{equation}
where the constraints of the boundary condition from Eq.~\ref{eqn2} have been imposed by using the Faddeev-Popov method~\cite{faddeev2018gauge}. Integrating over the dislocation degrees of freedom, we have 
\begin{equation}
\begin{aligned}
S[\phi] &= S_\text{ph}[\phi]+n_\text{dis} \sum_{n, \textbf{s}} C_\textbf{s}\Phi_{n\textbf{s}} \\
&-\frac{n_\text{dis}^2}{2 \beta } \sum_{n, \textbf{k}}\frac{\Omega_\textbf{k}}{\omega_n^2+ \Omega_\textbf{k}^2}\Phi_{n\textbf{k}}\Phi_{-n-\textbf{k}}
\label{eq-actiopmritive}
\end{aligned}
\end{equation}

To clearly see the quantum phase transition, without loss of generality we keep the leading contribution of the quadratic term, namely
\begin{equation}
\sum_{n,\textbf{s}}C_\textbf{s}\Phi_{n\textbf{s}} \approx -\sum_{n,\textbf{k}}\frac{\Delta}{\omega_\textbf{k}}\overline{\phi}_{n\textbf{k}}\phi_{n\textbf{k}}
\end{equation}
where a dynamic coupling constant $\Delta$ is introduced, arising from the particular form of the classical dislocation-phonon scattering amplitude $C_{\textbf{s}}$ and hence from the topological definition of the dislocation. The quartic term can further be encapsulated into an effective anharmonic coupling constant $g$:
\begin{equation}
\frac{1}{4!}g \equiv \frac{1}{4!}g_{4} - \frac{1}{2} \frac{A_\textbf{k}^2 \Omega_\textbf{k}}{\omega_n^2+ \Omega_\textbf{k}^2}
\end{equation}
where $g_{4}$ is the four-phonon anharmonic constant without dislocations. Notably, the phonon effective action, Eq.~\ref{eq-actiopmritive}, is then simplified as 
\begin{equation}
S[\phi]= S_{0}[\phi]+S_\text{int}[\phi]
\label{action}
\end{equation}
with  
\begin{equation}
\begin{aligned}
 S_{0}[\phi] &= \sum_{n\textbf{k}} \overline{\phi}_{n\textbf{k}} \left(-i\omega_{n} + \omega_{\textbf{k}} - \frac{n_\text{dis} \Delta}{\omega_{\textbf{k}}} \right) \phi_{n\textbf{k}} \\
S_\text{int}[\phi]&=\frac{g}{4!} \int_{0}^{\beta} d\tau \int d^{D} \textbf{R} \phi^{4} (\textbf{R}, \tau)
\end{aligned}
\end{equation}
where $D$ is the space dimensionality.

From Eq. \ref{action}, we obtain the non-interacting momentum-space propagator
\begin{equation}
D_0(\textbf{k},\omega_n) =\frac{1}{2} \left(\frac{1}{-i\omega_{n} + \omega_\textbf{k} - \frac{n_\text{dis} \Delta}{\omega_\textbf{k}}}\right)
\label{propagator}
\end{equation}
from which we immediately see the existence of a quantum critical point at zero temperature ($\omega_n=0$), where the critical phonon frequency $\omega^c$ at the mean-field level can be expressed as 
\begin{equation}
\omega^c=\sqrt{n_{\text{dis}}\Delta}. 
\end{equation}
At $\omega_\textbf{k}=\omega^c$, the divergence of the propagator in Eq.~\ref{propagator} confirms the existence of quantum criticality. Fig. \ref{figureone} schematically illustrates the two phonon phases, separated by the critical point $\omega^c$. The symmetric phase has $\langle\phi\rangle=0$, corresponding to conventional phonons; after the phase transition, a new phonon phase with broken symmetry is developed with $\langle\phi\rangle\neq0$, indicating the existence of net lattice displacement and the corresponding dynamically-induced dipole field, which is generally observed only in ferroelectric materials~\cite{lines2001principles}.

To explore the critical behavior beyond the mean field level,  we employ the momentum shell renormalization group (RG) analysis, in which we split the field appearing in Eq.~\ref{action} such that  $\phi_{n\textbf{k}}=\phi^s_{n\textbf{k}}+\phi^f_{n\textbf{k}}$, where $\phi^s_{n\textbf{k}}$ contains slow momentum components with $|\textbf{k}|< \Lambda/b$ and $b>1$, and $\phi^f_{n\textbf{k}}$ contains the fast momenta $\Lambda/b<|\textbf{k}|<\Lambda$. The ultraviolet (UV) cutoff is denoted by $\Lambda$, which can be approximated by a hypersphere of radius $\Lambda \approx \pi/a$ with  $a$ being the short length-scale cutoff. 

The RG calculation proceeds by integrating out the fast modes residing in the momentum shell $ \Lambda/b \leq |\textbf{k}| <\Lambda$, which results in a renormalized phonon effective action for the slow modes  $\phi^s_{n\textbf{k}}$. In the following rescaling step, the UV cutoff for the slow-mode effective action $\Lambda/b$ is restored to its original value $\Lambda$ by rescaling the fields and momenta, giving rise to RG equations for the dynamic coupling strength $\Delta$ and the anharmonic coupling constant $g$. In this work, we compute the RG equations up to one-loop order; in addition, in the interaction action $S_\text{int}[\phi^{f}+\phi^{s}]= \frac{g}{4!} \int_{0}^{\beta} d\tau \int d^{D}\textbf{R} ( \phi^{4,f} +4 \phi^{3,f} \phi^{s}+ 6\phi^{2,f} \phi^{2,s}+ 4\phi^{f} \phi^{3,s}+\phi^{4,s})$, we keep only the $\phi^{2,f}\phi^{2,s}$ term since it is the only one contributing to the one-loop RG analysis ~\cite{zinn1996quantum}. The relevant diagrams appearing in the renormalization of the coupling parameters $\Delta$ and $g$ are shown in Fig.~\ref{digrams}.  The first graph in Fig.~\ref{digrams} renormalizes the quadratic term of $\phi^s$ and hence renormalizes $\Delta$, while the second graph in  Fig.~\ref{digrams} renormalizes the quartic term in $\phi^s$ and hence renormalizes $g$. The internal lines are evaluated using the non-interacting propagator $D_0(\textbf{k},\omega_n)$ in Eq. \ref{propagator}. Collecting terms in this manner,  we obtain the renormalized phonon effective action in terms of $\phi^s$:

\begin{figure}
  \centering
  \includegraphics[width=8cm]{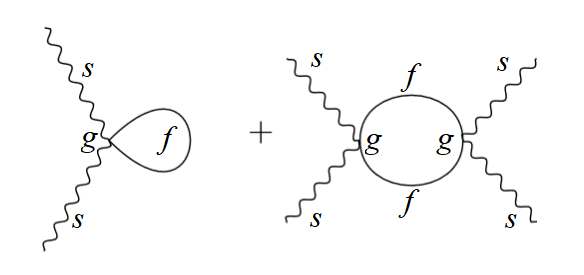}
\caption{Feynman diagrams to one-loop corrections of phonon field $\phi$. The external wavy lines represent $\phi^s$, while the internal lines $\phi^f$ propagators. The first graph renormalizes dynamic coupling $\Delta$, while the second renormalizes anharmonicity $g$.}
 \label{digrams}
   \end{figure}

\begin{equation}
\begin{aligned}
S_\text{eff}[\phi^s] = S[\phi^s] &+ K_1\sum_n \sum_{0<\textbf{k}<\Lambda/b}|\phi_{n\textbf{k}}^s|^2\\
&-K_2\int_0^\beta d\tau \int d^D\textbf{R} \phi^{4,s}(\textbf{R},\tau)
\end{aligned}
\end{equation}
with coefficients
\begin{equation}
\begin{aligned}
K_1&=\frac{3g}{4\beta} \sum_{n} \int_{\Lambda/b}^{\Lambda} \frac{d^{D}\textbf{k}}{(2\pi)^D} D_0 (\textbf{k},\omega_n),\\
 K_2&=\frac{3}{2\beta}\left(\frac{3g}{4!}\right)^2\sum_{n} \int_{\Lambda/b}^{\Lambda} \frac{d^{D}\textbf{k}}{(2\pi)^D} D_0^2 (\textbf{k},\omega_n). 
\end{aligned}
\end{equation}
The coefficients $K_1$ and $K_2$ can be calculated using the standard Matsubara summation method~\cite{kapusta2006finite}.

We now can immediately implement the rescaling step of the RG analysis to obtain the RG flow of the coupling constants. Taking into account the scaling dictated by $\textbf{k}\to b \textbf{k}$, $\omega_n \to b^z \omega_n$, $\omega_\textbf{k} \to b^{z_0} \omega_\textbf{k}$, $\phi_{n\textbf{k}}^s \to b^{y_\phi}\phi_{n\textbf{k}}^s$, $\tau \to b^{-z_0} \tau$ and $\phi(\textbf{R},\tau) \to b^{\frac{z_0+D}{2}+y_\phi}\phi(\textbf{R},\tau)$, and  transforming back to the original cutoff, we obtain the following RG flow equations: 
\begin{equation}
\begin{aligned}
 \frac{d\Delta}{dl} &= (2z_0-2)\Delta+ Ag\\
 \frac{dg}{dl} &= (z_{0}-D)g + Bg^2 \\
 \frac{d\beta}{dl} &= -z_{0}\beta
\end{aligned}
\label{eq-RGflowequa}
\end{equation}
where  $A$ and $B$  are coefficients which can be written as

\begin{equation}
\begin{split}
A&=\frac{1}{8}\frac{S_{D-1}}{(2\pi)^D}\Lambda^D (n_B(c\Lambda^{z_{0}})+1/2)\\
B&=\frac{3}{64}e^{\beta c\Lambda^{z_{0}}}n_B^2(c\Lambda^{z_{0}}) \frac{S_{D-1}}{(2\pi)^D}\Lambda^D.
\end{split}
\label{AB}
\end{equation}
Here, $n_B(a)=1/(e^{\beta a}-1)$ is the Bose distribution function, and $S_{D-1}$ denotes the area of a $(D-1)$-sphere. In defining the flow equations, we assumed a generic phonon dispersion $\omega_\textbf{k}=c|\textbf{k}|^{z_0}$. For example, for acoustic phonons, $\omega_\textbf{k}=c|\textbf{k}|$ with $z_0=1$, while for optical phonons, we have  $\omega_\textbf{k}=c$ with $z_0=0$. Therefore,  Eq.~\ref{AB} is suited for all types of phonons. Although the scaling equations are valid at any temperature, we consider two limits: a low temperature regime close to the quantum limit, which occurs near zero temperature and is dominated by the quantum fluctuations, and a high temperature regime where thermal fluctuations prevail over quantum fluctuations~\cite{fisher1988dilute}. 

\begin{figure}
  \centering
  \includegraphics[width=8cm]{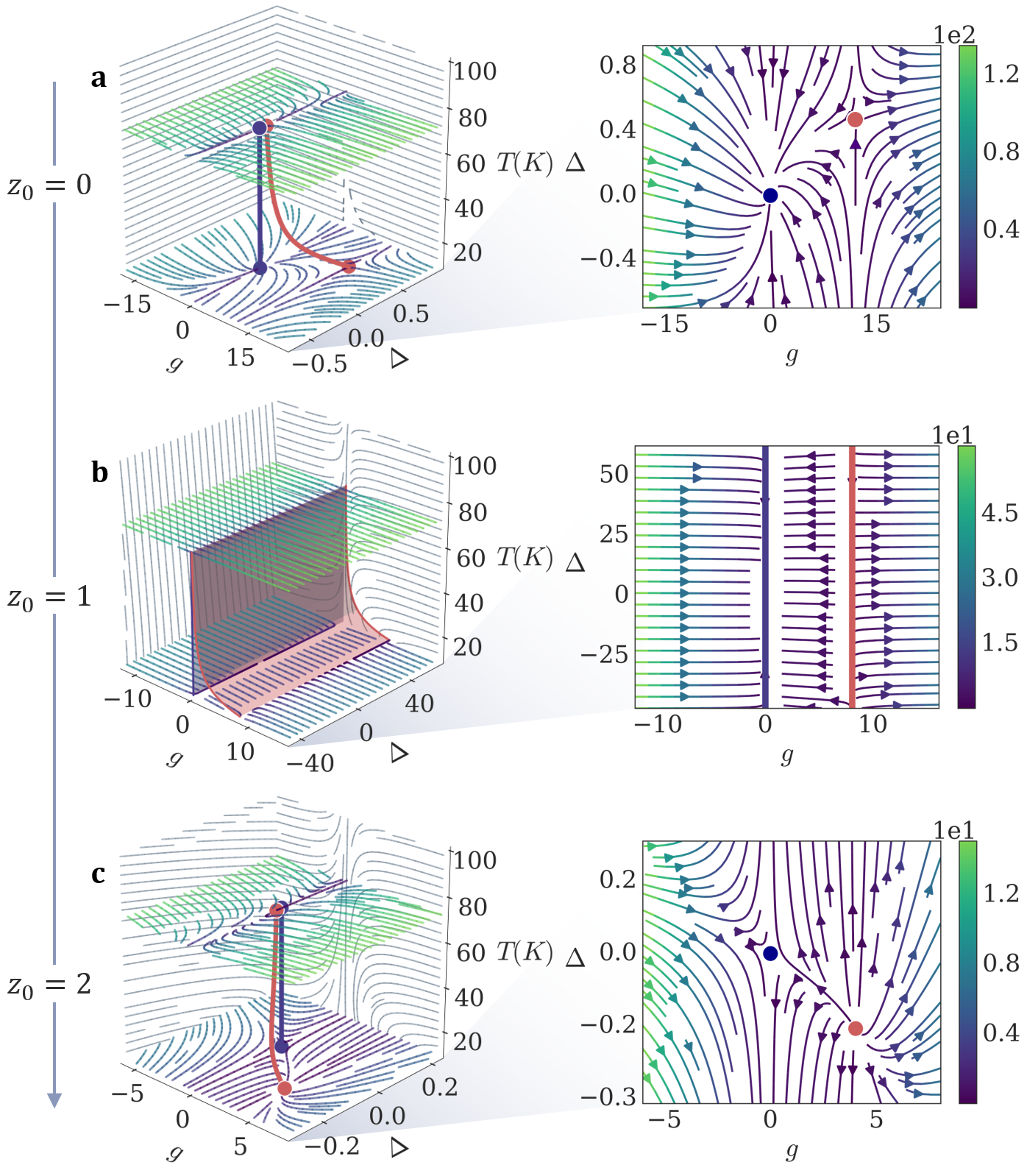}
  \caption{Renormalization group flows in the vicinity of fixed points for an anharmonic  phonon system. Here (a), (b) and (c) represent optical ($z_0=0$), acoustic ($z_0=1$) and quadratic ($z_0=2$) phonon cases at low and high temperatures. A realistic $c=1\times10^{-4}$ is being used.}
  \label{figurRGflow}
   \end{figure}

To analyze the RG equations given in Eq.~\ref{eq-RGflowequa}, we concentrate on the three-dimensional case ($D=3$)  for $z_0=0,1,2$ in the low and high temperature limits for which the RG flows are depicted in Fig.~\ref{figurRGflow}. Since the RG flow is computed at fixed $\Lambda$, we have conveniently set $\Lambda=1$ and will henceforth measure all lengths in units of $\Lambda^{-1}$. We obtain a trivial fixed point $g^*=\Delta^*=0$, representing the conventional phonon phase, and a non-trivial, critical fixed point $g_c^*=(z_0-D)/B$, $\Delta_c^* = Ag_c^*/(2-2z_0)$, representing the exotic phonon state. The divergence of $\Delta_c$ at $z_0 = 1$ indicates single-parameter scaling controlled solely by $g$ for acoustic phonons. For all three cases of $z_0$, the temperature dependence of the fixed points is shown in the left column of Fig.~\ref{figurRGflow}. As temperature increases, the Gaussian fixed points (blue dots) remain invariant while the quantum critical points (red dots) gradually approach the Gaussian fixed points, indicating negligible quantum fluctuations and a diminished quantum critical phase at elevated temperatures.

At all temperatures, there is an antisymmetry of the RG flow with respect to the $\Delta$ axis (Fig.~\ref{figurRGflow}). This is natural since the sign change of the coupling constant $g \to -g$ can be compensated by a phase shift of the field $\phi \to \phi e^{i\pi}$. Hence, it is sufficient to consider only the upper half-plane, $g\geq 0$. 
For optical phonons ($z_0=0$), $\beta$ in Eq.~\ref{eq-RGflowequa} does not flow, hence the temperature is kept constant in the RG flow. The analysis of stability shows that the fixed point at $g^*=\Delta^*=0$ (blue dot) is stable against thermal fluctuations, while the fixed point at $g_c=(z_0-D)/B$ (red dot) is unstable, as displayed in Fig.~\ref{figurRGflow}a. In addition, we observe $\Delta_c>0$ for optical phonons. Since $\Delta_c$ can be traced back to the scattering amplitude $C_s$, the condition $\Delta_c>0$ suggests that the QPT can take place when the dislocation scattering does not cause a large phase shift ($> \pi$) between incident and outgoing phonons, which is easily achieveable. 

For acoustic phonons ($z_0=1$) (see Fig.~\ref{figurRGflow}b), both the Gaussian and critical fixed points evolve into two straight ``fixed lines" controlled solely by $g$, and the usual stability of the Gaussian fixed point remains. On the contrary, the critical fixed line at $g_c$ is unstable. In particular, when $g<g_c$, the system will be driven into the stable phase at $g^*=0$; conversely, when $g>g_c$, the RG flow will be driven to $g \to \infty$, suggesting both the necessity to incorporate higher-order correction ~\cite{raposo1997quantum}, and the possibility of other novel phonon phases beyond the one-loop correction.
As for the quadratic case $z_0=2$, since $z_0-1>0$, according to Eq. \ref{eq-RGflowequa}, the Gaussian fixed point ($g^*,\Delta^*$) now becomes critical, while the non-trivial fixed point $(g_c^*,\Delta_c^*$) becomes unstable, as displayed in Fig.~\ref{figurRGflow}c. Overall, the co-existence of multiple phonon branches enables the possible realization of multiple phononic critical points in one system, which is extremely rare in an electronic system.

To understand  the influence of phonon quantum criticality on thermodynamic properties, we notice that the Green's function $D_0(\textbf{R},\tau)$ can be written in terms of Eq. \ref{propagator} as

\begin{equation}
D_0(\textbf{R},\tau)= \frac{1}{\beta}\sum_n \int \frac{d^D\textbf{k}}{(2\pi)^D}  D_{0}(\textbf{k},\omega_n) e^{i\textbf{k}\cdot \textbf{R}+i\omega_n\tau}
\label{proptrsd}
\end{equation}
From this point forward, we focus only on acoustic phonons since they dominate the thermal transport. Under the dilute-dislocation limit, we have $\omega_\textbf{k} - n_\text{dis}\Delta/\omega_\textbf{k} \approx \omega_\textbf{k}=c |\textbf{k}|$ where $c$ is  the sound velocity. At short range, we further expect the dominant contribution in the Fourier transform of  Eq.~\ref{proptrsd} to come from the regime $|\textbf{k} \cdot \textbf{R}|\approx 1$, so we assume $|\textbf{k}|>>T$ . Consequently, we have

\begin{equation}
D_0(\textbf{R},\tau)=\frac{\Gamma(D)}{(4\pi c^2)^{D/2}\Gamma(D/2)|\textbf{R}|^D |\tau|^D} 
\end{equation}
Therefore, in the critical regime, where the correlation function exhibits power-law behavior with a non-universal exponent $\eta$, our theory predicts $\eta=2$ even at mean field level. This is in sharp contrast to the conventional ferromagnetic $\phi^{4}$ theory, which has $\eta=0$~\cite{tauber2012renormalization}.  

Following the standard definition of specific heat~\cite{tauber2012renormalization}, we obtain the frequency-dependent specific heat 

\begin{equation}
C(\omega) \sim  \left(\frac{\omega-\omega^c}{\omega^c} \right)^{-\alpha}
\end{equation}
where the critical exponent $\alpha=1$ for acoustic phonons. This is also different from an conventional $\phi^{4}$ theory with mean field $\alpha=0$.

The lattice thermal conductivity can be written as~\cite{GchenPRB} 

\begin{equation}
k = \frac{1}{3}\int C(\omega) v_{\omega}^2\tau d\omega \propto \ln \left(|\omega-\omega^c| \right) 
\end{equation}
This leads to highly exotic observable effects. In particular, the thermal conductivity $k$ shows a logarithmic divergence near the critical point, determined by a critical frequency.

In this work, we demonstrated the existence of phonon quantum criticality driven by the competition between the intrinsic topology of a dislocation and the extrinsic inelastic phonon-dislocation anharmonic scattering. At $0$K, we found a simple expression of the critical point $\omega^c=\sqrt{n_{dis}\Delta}$. The resulted symmetry-broken phonon phase has a dynamically induced dipole field, and can be preserved at finite temperature as long as quantum fluctuations dominate thermal fluctuations. The scaling behavior is distinct from an electronic system, as the RG flow of coupling constants is highly dependent on the phonon dispersion. The resulting novel quantum critical phase raises a more general question, whether other exotic material phases can emerge as a result of competition between topological protection and a topology-breaking mechanism (here, inelastic scattering). Moreover, the dynamically-induced dipole field and phonon softening may be used to quantify defect-
density in semiconductors with high precision \cite{Irmer}. Most importantly, at the critical point, one single phonon frequency dominates the density of states and thereby resembles a fermionic system by forming a ``fermi-surface-like" structure in a bosonic system. Possible observation of this effect, such as in a defective noble-gas solid with strong quantum fluctuations, may have huge impact for thermal transport by converting the broad-band phonon -- a purely bosonic system -- into a narrow-band one. 

\begin{acknowledgments} 
We thank Ian Fisher, Alan Tennant and Gang Chen for helpful discussions. R.P.P. acknowledges the support from La Associacion Civil M\'exico en Movimiento and Becalos Foundation. N.A. acknowledges the support of the National Science Foundation Graduate Research Fellowship Program under Grant No. 1122374. Y.T., Z.D and T.-H. L. thank support from Department of Defense, Defense Advanced Research Projects Agency (DARPA) Materials for Transduction (MATRIX) program, under Grant HR0011-16-2-0041. 
\end{acknowledgments}

R.P.P. and N.A. contributed equally to this work.

\bibliographystyle{apsrev.bst}
\bibliography{main}
%\pagebreak
%\onecolumngrid
%\appendix
%\setcounter{figure}{0}
%\renewcommand{\thefigure}{A\arabic{figure}}
%\input{supplementary_text}

\end{document}

%% file: authors.tex
\author{Ricardo Pablo-Pedro}
%\thanks{These authors contributed equally to this work.}
\email{ripablo@mit.edu}
\affiliation{Energy Nano Group, MIT, Cambridge, Massachusetts 02139, USA}
\affiliation{Department of Nuclear Science and Engineering, MIT, Cambridge, Massachusetts 02139, USA}
\author{Nina Andrejevic}
%\thanks{These authors contributed equally to this work.}
\affiliation{Energy Nano Group, MIT, Cambridge, Massachusetts 02139, USA}
\affiliation{Department of Materials Science and Engineering, MIT, Cambridge, Massachusetts 02139, USA}
\author{Yoichiro Tsurimaki}
\affiliation{Department of Mechanical Engineering, MIT, Cambridge, Massachusetts 02139, USA}
\author{Zhiwei Ding}
\affiliation{Department of Materials Science and Engineering, MIT, Cambridge, Massachusetts 02139, USA}
\affiliation{Department of Mechanical Engineering, MIT, Cambridge, Massachusetts 02139, USA}
\author{Te-Huan Liu}
\affiliation{Department of Mechanical Engineering, MIT, Cambridge, Massachusetts 02139, USA}
\author{Gerald D Mahan}
\affiliation{Department of Mechanical Engineering, MIT, Cambridge, Massachusetts 02139, USA}
\affiliation{Department of Physics, The Pennsylvania State University, University Park, PA 16802, USA}
\author{Shengxi Huang}
\affiliation{Department of Electrical Engineering, The Pennsylvania State University, University Park 16802, USA}
\author{Mingda Li}
\email{mingda@mit.edu}
\affiliation{Energy Nano Group, MIT, Cambridge, Massachusetts 02139, USA}
\affiliation{Department of Nuclear Science and Engineering, MIT, Cambridge, Massachusetts 02139, USA}